\documentclass[conference]{IEEEtran}

\usepackage{cite}      
\usepackage{graphicx}  
\usepackage{psfrag}                            
\usepackage{subfigure} 
\usepackage{url}      
\usepackage{stfloats}                          
\usepackage{amsmath}   
\usepackage{array}
\usepackage{fancyhdr}

\begin{document}

\title{\vspace{50.0pt} An Economic Model of Coupled Exponential Maps}

\author{\authorblockN{Ricardo L\'opez-Ruiz}
\authorblockA{Facultad de Ciencias \\
Universidad de Zaragoza \\ E-50009 Zaragoza, Spain\\
Email: rilopez@unizar.es}
\and
\authorblockN{Javier Gonz\'{a}lez-Est\'{e}vez}
\authorblockA{Laboratorio de F\'{\i}sica Aplicada \\ y Computacional, U.N.E.T.\\
San Crist\'{o}bal, Venezuela\\
Email: jgonzale@unet.edu.ve}
\and
\authorblockN{Mario G. Cosenza}
\authorblockA{Centro de F\'{\i}sica Fundamental \\ 
Universidad de Los Andes \\ M\'{e}rida 5251, Venezuela \\
Email: mcosenza@ula.ve}
\and
\authorblockN{Juan R. S\'{a}nchez}
\authorblockA{Facultad de Ingenier\'{\i}a \\
Univ. Nac. de Mar del Plata, \\ 7600-Mar del Plata, Argentina \\
Email: jsanchez@fi.mdp.edu.ar}}


\maketitle
\thispagestyle{fancy}
\markboth{\scriptsize IEEE Workshop on Nonlinear Maps and Applications (NOMA'07)}
{\scriptsize LATTIS-INSA, Toulouse University, France, Dec. 13-14, 2007}


\begin{abstract}

In this work, an ensemble of economic interacting agents is considered. 
The agents are arranged in a linear array where only local couplings are allowed.
The deterministic dynamics of each agent is given by a map. This map is expressed by 
two factors. The first one is a linear term that models the expansion of the agent's economy
and that is controlled by the {\it growth capacity parameter}.  
The second one is an inhibition exponential term that is regulated by the {\it local environmental pressure}.
Depending on the parameter setting, the system can display Pareto or Boltzmann-Gibbs 
behavior in the asymptotic dynamical regime. 
The regions of parameter space where the system exhibits one of these two statistical behaviors
are delimited. Other properties of the system, such as the mean wealth,
the standard deviation and the Gini coefficient, are also calculated.

\end{abstract}

\IEEEpeerreviewmaketitle

\section{Introduction}

Nowadays it is well established that the wealth distribution in western societies 
presents essentially two phases. This means that the whole society can be split in 
two disjoint parts in which the richness distribution is different in each of them. Thus,
Dragulescu and Yakovenko \cite{yakovenko2001} have found, over real economic data from 
UK and USA societies, that one phase presents an exponential (Boltzmann-Gibbs $\rightarrow$ BG)
distribution which covers about $95\%$ of individuals, those with low and medium incomes,
and that the other phase, which is integrated by the high incomes, i.e., the $5\%$ 
of individuals, shows a power (Pareto) law.  
Different dynamical mechanisms for the interaction among agents in multi-agent 
economic models have been proposed in the literature in order to reproduce
these two types of statistical behavior.

In the Dragulescu and Yakovenko \cite{yakovenko2000} model,
a set of $N$ economic agents exchange their own amount of money $u_i$, with $i=1,2,\cdots,N$, 
under random binary $(i,j)$ interactions, $(u_i,u_j)\rightarrow (u_i',u_j')$, 
by the following exchange rule:
\begin{eqnarray}
u'_i & = & \epsilon (u_i+u_j), \\
u'_j & = & \bar\epsilon(u_i+u_j), 
\end{eqnarray}
with $\bar\epsilon=(1-\epsilon)$, and $\epsilon$ a random number in the interval $(0,1)$.
Starting the system from an initial state with equity among all the agents, 
it evolves towards an asymptotic dynamical
state where the richness follows a BG distribution \cite{lopezruiz2007}.
This model is equivalent to the saving propensity model introduced by 
Chakraborti and Chakrabarti \cite{chakraborti2000} for the particular case 
in which the agents do not save any fraction of money before carrying out each transaction. 
A similar model was introduced by Angle \cite{angle2006}.
In this case, the random pair $(i,j)$ of economic agents, $(u_i,u_j)$, 
exchanges money under the rule:
\begin{eqnarray}
u'_i & = & u_i-\Delta u, \\
u'_j & = & u_j+\Delta u, 
\end{eqnarray}
where
\begin{equation}
\Delta u=\epsilon\omega u_i,
\label{eq-delta}
\end{equation}
with $\epsilon$ a random number in the interval $(0,1)$. The exchange parameter, $\omega$, 
with $0<\omega< 1$, represents the maximum fraction of wealth lost by the agent $u_i$ in the transaction. 
When $\omega=0.75$, the exponential distribution of the wealth is also found for this model.  

In order to recover a power law behavior, these models can be reformulated in a nonhomogeneous way.
These are models in which each agent $i$ can have assigned by some random function 
a different value $\omega_i$ of the exchange parameter $\omega$ in the Angle model \cite{angle2006} 
or a different saving propensity in the Chakraborti and Chakrabarti model \cite{chakraborti2000}. 
Under quite general conditions, robust Pareto distributions have been found 
at large $u$-values in these reformulations \cite{chakraborti2000,angle2006}.

Randomness is an essential ingredient in all the former models.
Thus, agents interact by pairs chosen at random, and these pairs 
exchange a random quantity of money in each transaction. Moreover, the transition from the
BG to the Pareto behavior requires the change of the structural properties
of the system. It can be reached, for instance, by introducing a strong inhomogeneity that breaks
the initial indistinguishability among the units of the ensemble. From a practical point of view, 
this could seem an unrealistic approach since we need to conform very different setups
in order to mimic the two statistical behaviors. On the other hand, interactions among 
the economic or social agents of a real collectivity are not fully random. In fact, 
they are driving in the majority of transactions by some kind of mutual interest or rational forces. 
Hence, it would be useful to dispose of a multi-agent model with the ability of displaying 
BG and Pareto behaviors emerging from an asymptotic dynamics  
where determinism would play some role in the evolution of the system.
Now, we proceed to show with some detail the properties of one of 
these models\cite{sanchez2007} recently introduced in the literature.

\section{The Model}

The model \cite{sanchez2007} consists of a linear array of $N$ interacting agents
with periodic boundary conditions.  Each agent, which can represent a company, country 
or other economic entity, is identified by an index $i$, with $i=1 \cdots N$ 
and $N$ being the system size. Its actual state is characterized by a real number, $x_i$, 
denoting the {\it strength, wealth or richness} of the agent, with $x_i\in [0,\infty)$. 
The system evolves in time synchronously, and only interactions among nearest-neighbors
are allowed. Thus, the state of the agent, $x_i^{t+1}$, at time $t+1$ is given by the product of 
two terms at the precedent time $t$; 
the {\it natural growth} of the agent, $r_ix_i^{t}$, with its own {\it growth capacity parameter},
$r_i$, and a {\it control term} that limits this growth with respect 
to the local field $\Psi_i^t=\frac{1}{2}(x_{i-1}^{t}+x_{i+1}^{t})$ through a 
negative exponential function with {\it local environmental pressure} $a_i$:

\begin{equation}
x_i^{t+1} = r_i\:x_i^{t}\: \exp(-\mid x_i^{t}-a_i\Psi_i^t\mid).
\label{eq:system}
\end{equation}

The parameter $r_i$ represents the {\it capacity} of the agent to become richer 
and the parameter $a_i$ describes some kind of local {\it pressure} \cite{ausloos2003} 
that saturates its exponential growth. 
This means that the largest possibility of growth for the agent is obtained when 
$x_i\simeq a_i\Psi_i^t$, i.e., when the agent has reached some kind of adaptation 
to the local environment. In this note, for the sake of simplicity, we concentrate 
our interest in a homogeneous system with a constant capacity $r$ and a constant 
selection pressure $a$ for the whole array of sites. 
 
If all the agents start with the same wealth, the index $i$ can be omitted, 
$x_i^t=x^t$ and $\Psi_i^t=x^t$, and the global evolution reduces to the following map,
\begin{equation}
x^{t+1}=r x^{t} \exp(-\mid 1-a\mid x^{t}).
\end{equation}

The above map can be easily analyzed by standard techniques. In fact,
the parameter $a$ could be removed by doing the change of variable
$y^t= \mid 1-a\mid x_t$, and 
obtaining the generic map $y^{t+1}=r y^{t} \exp(-y^{t})$.
For $r<1$ the system relaxes to zero and for $r>1$ 
the dynamics can be self-sustained deriving toward 
different types of attractors. It displays all kind of bifurcations known for 
this type of maps \cite{schuster1984}, except, evidently, for the singular case $a=1$. 
For instance, when $r>1$ the fixed point is $x_0=\log r/\mid 1-a\mid$.
This point becomes unstable by a flip bifurcation for $r=e^2$. For increasing $r$ 
the whole period doubling cascade and other complex dynamical behavior are obtained. 
However, it can be shown that such evolving uniform states are unstable. When a 
perturbation is introduced in the initial uniform state or, in general, when the 
initial condition is a completely random one, the asymptotic dynamical state of 
the system is found to be more complex.

\section{Some Results}

We perform the statistical study of the system in the parameter space $(a,r)$.
For all the simulations, the system size is $N=10^5$ and the initial conditions in the array
are random values in the interval $(1,100)$. Also, a transient of $10^4$ iterations is completed 
before arriving to the asymptotic state where all the measurements are done. At this point,
if necessary, the average is done over the next $100$ iterations after the transient, and 
this result is newly averaged with the same process over $100$ different realizations of the system.
Following this method, different statistical quantities,
such as the mean wealth, the standard deviation and the Gini coefficient, 
have been calculated on the model, and they will be shown in the communication 
to be presented in NOMA'07.

As an example, we advance here the possibility of existence of the two statistical behaviors 
in which we are interested in. 
Hence, if the number of agents, $P(x)$, with a wealth $x$ is represented
in a semi-log plot, a BG distribution is found for the values $a=0.6$ and $r=4$
(Fig. 1a, where values of maps have been taken at time $t=10^4$).
The exponent $\mu$ of this distribution, $P(x)\sim e^{-\mu x}$,
is $\mu=0.26$. A thermodynamical simile can be done for this type of behavior by defining
a kind of `temperature', $h=1/\mu$, that is related with the mean wealth of each agent
in the ensemble. In this particular case, $h=3.84$. 

\hskip 1cm

\begin{figure}[h]
\centerline{\includegraphics[width=9cm]{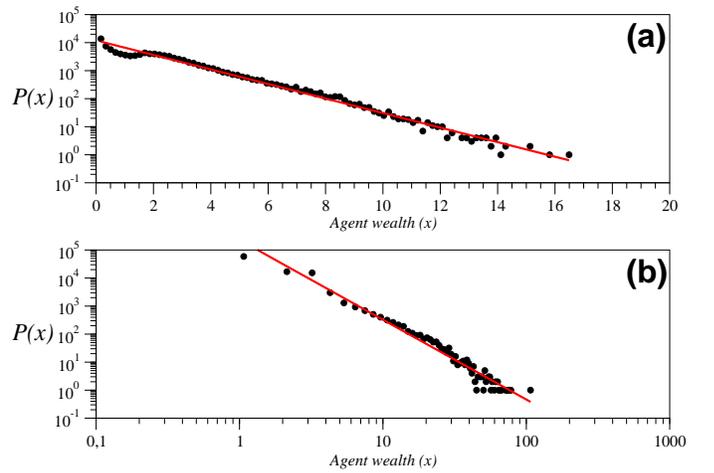}}
\caption{{(\bf 1a):} Semi-log plot of the number of agents, $P(x)$, with a wealth $x$
for $a=0.6$ and $r=4$ in the asymptotic regime at time $t=10^4$. The scaling
of this exponential behavior is $0.26$. 
{(\bf 1b):} Log-log plot of  $P(x)$ for $a=0.92$ and $r=8$ 
at time $t=10^4$. The exponent of this power law behavior is $2.84$.} 
\label{fig1}
\end{figure}

Also, a power law behavior in the statistical behavior of system ($\ref{eq:system}$) 
is possible, for instance, for $a=0.92$ and $r=8$.
If the number of agents, $P(x)$, with a wealth $x$ is now represented
in a log-log plot, a Pareto distribution is unmasked \cite{sanchez2007} 
(Fig. 1b, where values of maps have also been
taken at time $t=10^4$). The exponent $\alpha$ of this distribution, $P(x)\sim x^{-\alpha}$,
is in this case $\alpha=2.84$, which is in clear agreement with the exponents derived from
real economic data \cite{yakovenko2001,levy1997,souma2001}. 
Thus, the exponent $\bar\alpha$ found in the cumulative probability 
distribution of incomes, whose scaling behaves as $x^{-\bar\alpha}\sim x^{-\alpha+1}$, 
is about $2.1$ for the UK and $1.7$ for the US economies \cite{yakovenko2001}. 
Pareto himself proposed a value of $1.5$, Levy and Solomon \cite{levy1997}
found a value of $1.36$ for the distribution of wealth in the Forbes $400$ and
Souma \cite{souma2001} found $2.05$ for the high income distribution in Japan.

\section{Conclusion}

Let us conclude this note by remarking that our present model (\ref{eq:system})
shows, at least, two interesting statistical behaviors, 
those of Boltzmann-Gibbs and Pareto types, in its asymptotic state.
The procedure to get these two regimes does not require of any structural change in the 
system. Only, it is necessary to tune some adequate pair of values of the external parameters
$(a,r)$ for which the system displays one of those types of behavior. This property could be 
an advantage respect to other models in the literature that need to perform a strong change
in the conformation of the initial ensemble in order to present one of the two behaviors, 
exponential or power law.
Other relevant property of model (\ref{eq:system}) is its complete determinism. There is no
any kind of random ingredient in its evolution. Of course, this does not forbid in any way 
that the asymptotic state of the system can show some degree of spatio-temporal complexity,
including very large fluctuations of the agent's wealth. 
As indicated by Yakovenko in his last review on this subject \cite{yakovenko2007},
let us finish by saying that it seems well proved that western societies consist of two
different classes characterized by different distribution functions. However, the most part
of theoretical models on this subject do not produce two classes, although they do produce
broad distributions. We hope that the model here studied can be used in the next future 
as a tool in the agent-based theory capable of simulating two economic classes populations.

\end{document}